\documentclass[11pt]{article}
\usepackage[english]{babel}
\usepackage{amsmath,amssymb,amsbsy,amstext,amsthm,booktabs,cite,color,simplewick,exscale,relsize,slashed,graphicx,amsfonts,upgreek}
\usepackage{multirow,dcolumn,bm,enumerate,url,hyperref}

\usepackage{url}
\usepackage{comment}
\usepackage{graphicx}
\usepackage{caption}
\usepackage{subcaption}
\usepackage{amssymb}
\usepackage{gensymb}
\usepackage{cleveref}
\usepackage{siunitx}
\usepackage{multirow}
\date{}

\title{\bf Etching Plastic Searches for Dark Matter}

\author{Amit Bhoonah$^\eth$, Joseph Bramante$^{\dagger \eth}$, Brian Courtman$^\eth$, Ningqiang Song$^{\dagger \eth}$\\
{\small $^\eth$ The Arthur B. McDonald Canadian Astroparticle Physics Research Institute and} \\ {\small Department of Physics, Engineering Physics, and Astronomy,} \\ {\small Queen's University, Kingston, Ontario, K7L 2S8, Canada}\\
{\small $^\dagger$Perimeter Institute for Theoretical Physics, Waterloo, Ontario, N2L 2Y5, Canada}}

\begin{document}
\maketitle

\begin{abstract}
Large panels of etched plastic, situated aboard the Skylab Space Station and inside the Ohya quarry near Tokyo, have been used to set limits on fluxes of cosmogenic particles. These plastic particle track detectors also provide the best sensitivity for some heavy dark matter that interacts strongly with nuclei. We revisit prior dark matter bounds from Skylab, and incorporate geometry-dependent thresholds, a halo velocity distribution, and a complete accounting of observed through-going particle fluxes. These considerations reduce the Skylab bound's mass range by a few orders of magnitude. However, a new analysis of Ohya data covers a portion of the prior Skylab bound, and excludes dark matter masses up to the Planck mass. Prospects for future etched plastic dark matter searches are discussed.
\end{abstract}


\section{Introduction}\label{sec:intro}

Dark matter (DM) is critical to the formation of the early universe and the structure of galaxies, and determining dark matter's mass and couplings to visible matter is a high priority of modern science. While many searches for dark matter focus on finding a weakly interacting massive particles (WIMPs), with a mass adjacent by a few orders of magnitude to Standard Model particles, it has been known for decades that heavier and more strongly interacting dark matter could make up the cold material that forms the bulk of matter in our universe. In particular, it was well appreciated that dark matter could be a nuclear material, with a large mass and nuclear scattering cross section \cite{Witten:1984rs,Farhi:1984qu,DeRujula:1984axn,Goodman:1984dc,Drukier:1986tm}. Some ``nuclearite'' models have since been excluded, but essentially the same considerations apply to models of heavy composite dark matter \cite{Nussinov:1985xr,Bagnasco:1993st,Alves:2009nf,Kribs:2009fy,Lee:2013bua,Krnjaic:2014xza,Detmold:2014qqa,Jacobs:2014yca,Wise:2014jva,Wise:2014ola,Hardy:2014mqa,Hardy:2015boa,Gresham:2017zqi,Gresham:2017cvl,Bramante:2018qbc,Gresham:2018anj,Bramante:2018tos,Ibe:2018juk,Grabowska:2018lnd,Coskuner:2018are,Bai:2018dxf,Digman:2019wdm,Bai:2019ogh,Bramante:2019yss,Bhoonah:2020dzs,Clark:2020mna,Acevedo:2020avd}. In this paper we will consider the detection of strongly interacting massive particle dark matter (SIMPs) using prior data collected by plastic etch detectors searching for cosmogenic particles. 

Unlike WIMPs, SIMPs often have such a large interaction cross section with normal matter that their energy is depleted along their path to a detector. As a consequence, after repeatedly scattering while traveling through the Earth's atmosphere and crust, SIMPs are slowed and the kinetic energy they can impart to electrons and nuclei in underground detectors is reduced. If the SIMP cross section is large enough, its underground interactions will not overcome the energy threshold of underground dark matter searches. This makes above-ground searches for SIMPs beneficial, although in practice this results in additional particle interaction backgrounds from cosmic rays. On the other hand, because SIMPs interact very strongly, comparatively simple detectors are capable of detecting them. 

In this work we will study bounds on dark matter interactions with nuclei, obtained from prior searches for cosmogenic particles in etched plastic track detectors. We will be particularly focused on plastic track detectors employed aboard the Skylab space station to search for cosmic rays \cite{Skylab} and underground in the Ohya quarry to look for monopoles, nuclearites, and exotic particles \cite{Orito:1990ny}. When a particle passes through plastic polymer, its interactions with nuclei and electrons can break the plastic's molecular bonds binding them together. The resulting microscopic damage can be enlarged and observed by treating the plastic material with acid. After acid treatment, the damage paths can be observed as holes in the plastic -- in the case of many plastic layers, the orientation of consecutive holes left by a particle's passage can be used to validate that particle's path and even determine its energy as it traveled through the plastic \cite{Durrani:1987zv,fleischer1975nuclear}. However, plastic etch detectors require large particle energy deposition for detection, around $\sim$ GeV/cm, compared to underground dark matter experiments, which can require as little as a single $\sim$eV energy deposition for detection. Compared to sensitive underground dark matter search experiments, etched plastic polymer detectors are relatively easy to assemble and maintain, and as a consequence the scale and mass reach of these detectors achieved in the early 1990s still exceeds the scale of modern underground dark matter search experiments, albeit for sensitivity to much larger dark matter cross sections than those being searched for underground. 

Etched plastic track detectors have been used previously to constrain strongly interacting dark matter. Reference~\cite{Starkman:1990nj} analyzed Skylab's plastic track data \cite{Skylab} to bound dark matter's spin-independent interactions with nucleons. This bound was re-derived in \cite{McGuire:1994pq} and later adapted for large composite dark matter models in \cite{Jacobs:2014yca}. Ohya quarry data \cite{Orito:1990ny} was also considered in Reference~\cite{McGuire:1994pq}. After these derivations, the Skylab nucleon cross section bound that has appeared most in the literature was first presented in \cite{Wandelt:2000ad,Erickcek:2007jv}, before subsequently being reproduced in many papers that considered heavy strongly interacting dark matter, $e.g.$ \cite{Mack:2007xj,Bhoonah:2018gjb,Digman:2019wdm,Bramante:2019fhi,Garani:2019rcb,Bai:2020ttp,Bhoonah:2020dzs,Cappiello:2020lbk,Acevedo:2020gro,Acevedo:2020avd}. Our purpose in this note will be to provide a detailed bound on dark matter's interactions using plastic track detectors, by incorporating a realistic dark matter flux, the geometry and composition of the detector and overburden, and an accurate energy threshold in the case of \cite{Orito:1990ny}. We will find that using a Maxwell-Boltzmann (MB) dark matter halo velocity distribution (instead of a fixed DM velocity $\sim 10^{-3}{\rm c}$) and carefully accounting for background events, appreciably changes plastic etch bounds, lessening the previously reported Skylab mass reach by orders of magnitude, while somewhat improving low cross section sensitivity.

The remainder of this note is organized as follows. In Section \ref{sec:dm}, we detail the distribution and detection of dark matter at plastic track experiments. In Section \ref{sec:bounds} we provide a detailed set of bounds on dark matter's cross section with nuclei using data from Skylab and Ohya \cite{Skylab,Orito:1990ny}. In Section \ref{sec:conc} we conclude. Appendix \ref{app:etch} provides a review of plastic track detectors and thresholds.

\section{Dark matter at etched plastic track detectors} 
\label{sec:dm}

As it traverses a polymer, dark matter can scatter against nuclei, breaking apart molecules in a fashion that can later be detected. In this work, we will assume two forms for dark matter interactions with nuclei. The first is a spin-independent nuclear scattering cross section, whereby dark matter interacts coherently with all nucleons in a nucleus. The nuclear cross section in this case is customarily defined in terms of a per nucleon cross section $\sigma_{\chi n}$ as
\begin{align} \label{eq:sxn}
    \sigma_{\chi A} &= A^{2}\frac{\mu^{2}_{\chi A}}{\mu^{2}_{\chi n}}\sigma_{\chi n}, &({\rm spin-independent})
\end{align}
where $A$ is the number of nucleons in a nucleus, $\mu_{\chi A}$ is the DM-nucleus reduced mass and $\mu_{\chi n}$ the DM-nucleon reduced mass. Here we have omitted nuclear form factors \cite{Lewin:1995rx}, as they will have a subdominant effect on our treatment.

Our second model encompasses the case where that dark matter interacts with the nucleus through a monotypical contact interaction. This is appropriate if the dark matter is much larger than the nucleus, and opaque to nuclei \cite{Jacobs:2014yca,Digman:2019wdm,Cappiello:2020lbk}. $I.e.$ if dark matter's radius is much larger than the nuclear radius, and all incident nuclei scatter, then the cross section is independent of the size or constitution of the nuclei. In this case for a contact cross section we will use
\begin{align}\label{eq:sc}
    \sigma_{C}, & &({\rm contact})
\end{align}
where this also assumes that the DM composite's structure does not affect scattering.

In terms of these quantities, the energy imparted to nuclei by dark matter with mass $m_{\chi}$ passing through a background of nuclei with number density $n_A$ is
\begin{align}\label{eq:dEdx}
   \frac{dE}{dx} = \frac{2 E}{m_{\chi}}\sum_{A}\frac{\mu^{2}_{\chi A}}{m_{A}}n_{A}\sigma_{\chi A},
   \end{align}
where the summation includes all background nuclei with $A$ nucleons, and $E = m_{\chi}v^{2}/2$ is the dark matter kinetic energy. The above expression applies to the spin-independent case. Here and in all that follows, equivalent expressions for $\sigma_C$ are obtained by substituting $\sigma_{\chi A} \rightarrow \sigma_C$.

Note that in Eq.~\eqref{eq:dEdx} we have averaged over elastic scattering angles. An averaged energy loss will be appropriate here, since we consider SIMPs that scatter many times while traversing detector and shielding material. Integrating this equation, we obtain an energy loss rate for a dark matter particle traversing a material over distance $x$,
\begin{align}\label{eq:EnergyLoss}
    E\left(x\right) = E\left(0\right)\,\exp\left[-2\frac{x}{m_{\chi}} \sum_{A}n_{A}\frac{\mu^{2}_{\chi A}}{m_{A}}\sigma_{\chi A}\right]
\end{align}
where $E\left(0\right) = m_{\chi}v^{2}/2$ is the initial energy of the dark matter particle. In practice, we will want to consider the dark matter's energy after traveling fully through an overburden length $x_{O}$ and detector length $x_D$,
\begin{align}\label{eq:EnergyLossTotal}
    E\left(x_{O} + x_{D}\right) = E(0)\,\, \exp\left[\frac{-2}{m_\chi} \left( x_O \sum_{A\subset O}n_{A}\frac{\mu^{2}_{\chi A}}{m_{A}}\sigma_{\chi A} +x_D \sum_{A \subset D}n_{A}\frac{\mu^{2}_{\chi A}}{m_{A}}\sigma_{\chi A} \right) \right]
\end{align}
where the sums are taken over nuclei $A$ in overburden material $O$ and detector material $D$.  

It remains to determine the velocity and density distribution of dark matter, which is an important input to determine the speed and frequency of dark matter particles incident on plastic track detectors. We will adopt a standard \cite{Lewin:1995rx} flux-normalized Maxwell-Boltzmann distribution with $ v_{0}$ the Milky Way's circular speed and $\vec v_E$ the earth's velocity in galactic rest frame,
\begin{align}\label{eq:BoltzDist}
   f\left(\vec v,\vec v_{E}\right)\,  d^{3} v \ = \frac{2 \pi}{\mathcal{N}}   \, v^{3} \,\exp\left(-\frac{\tilde v^2 }{v^{2}_{0}}\right) \Theta\left(v_{esc} -\tilde v \right)\,\, \sin \, \theta \, d\theta \, dv
\end{align}
where
\begin{align}
   \tilde v^2 \equiv v^2 + v_E^2 + vv_E\,\cos \,\theta
\end{align}
is the velocity of the dark matter particle in the rest frame of the galaxy,
$\mathcal{N}$ is a normalization constant such that $\int  f\left(\vec v,\vec v_{E}\right)\,d^{3} v  = 1$, the Heaviside theta function $\Theta$ enforces a cutoff such that dark matter's speed does not exceed the galaxy's escape speed $v_{esc}$ at Earth's position, and the factor of $2\pi$ indicates the assumption of no azimuthal angular dependence. We use the following speeds, where in an aim to be conservative in setting bounds we take the $1 \sigma$ lowered value for each of these measured quantities. For the escape speed from the Milky Way galaxy we take $v_{esc}$ = 503 km\,s$^{-1}$ \cite{2019MNRAS.485.3514D}, for the Milky Way circular velocity at the Sun's position $v_{0}$ = 222 km\,s$^{-1}$ \cite{Eilers2018} (applying their $3\%$ systematic uncertainty), and for the Sun's relative motion we take $v_{E}$ = 232 km\,s$^{-1}$ \cite{2010MNRAS.403.1829S}. For additional discussion, especially of the last quantity, see \cite{2020arXiv201202169B,2012ApJ...759..131B,Wegg2019,Sukanya}.

Now we are ready to specify the flux of dark matter at plastic track detectors. With a flux-normalized velocity distribution, the average velocity for incident dark matter is 
\begin{align}
    v_{ave} = \int_0^{\infty} v \, f\left(\vec v,\vec v_{E}\right) d^{3} v,
\end{align}
where it is important to note that $f\left(\vec v,\vec v_{E}\right)$ includes a $\Theta$ function cutting off the escape velocity, meaning the integral is truncated well before $v=\infty$.
The flux (in one direction) across a planar detector with area $A$ is then
\begin{align}\label{eq:flux}
    \Phi_{\chi} = \frac{\rho_{\chi}}{m_{\chi}} v_{ave}\, A \,f_g
\end{align}
where $\rho_{\chi} = 0.3 ~{\rm GeV/cm^3}$ is the dark matter energy density \cite{Eilers2018} and $f_g =\pi (1- \cos^2\theta_D)/(4\pi)$ 
is a geometric flux acceptance factor appropriate for a planar detector that accepts particles inside a zenith angle $\theta_D$, where $\theta_D = 0$ is the angle normal to the plane. In order to produce accurate and conservative bounds, we use Eq.~\eqref{eq:flux} to determine the DM flux. This means we only consider the DM flux that enters the planar detector from above -- we do this because Skylab and Ohya only provided detailed information about the material above their detectors.

We are now prepared to discuss dark matter searches at planar plastic track experiments. Both experiments we consider in the next section will have a planar design, and each observed $N_c > 10$ candidate events (most likely cosmic rays) that we will treat here as indistinguishable from SIMPs to conservatively set bounds on a potential dark matter flux. To set bounds on dark matter-nuclear interactions using plastic track detectors, we must require that more dark matter than $N_c$ would have passed through the material above the detector and the detector itself, with an energy large enough to leave observable polymer damage during the entire passage, $cf.$ Equation \eqref{eq:EnergyLossTotal}. Using the formulae derived in this section, we find it is possible to set plastic track bounds on dark matter detectors in two steps. 

\begin{enumerate}
    \item The maximum dark matter mass an experiment is sensitive to is found by matching the number of events observed at the plastic detector, $N_c$, to the expected dark matter flux over time $t$, $N_c = \Phi_\chi t$,
    \begin{align}\label{eq:maxmass}
        m_\chi^{\rm max} = \frac{\rho_\chi \, v_{ave}\, A \,f_g \, t}{N_c}.
    \end{align}
    \item Plastic track detector damage energy thresholds will favor detection of the fastest-moving SIMPs in the MB distribution $cf.$ Eq.~\eqref{eq:BoltzDist}. As mentioned above, to remain conservative we have selected $1 \sigma$ slow values for our velocity distribution parameters. In addition, for each dark matter mass we determine a slow initial velocity $v_i$, defined by 
    \begin{align}
        N_c = \Phi_\chi t \int_{v_i}^{\infty}f\left(\vec v,\vec v_{E}\right) d^{3} v 
    \end{align}
    For the calculations that follow we use $v_i$, which is the slowest speed of the high velocity portion of dark matter's MB distribution required to produce $N_c$ events in the detector over time $t$. Then substituting Eq.~\eqref{eq:EnergyLossTotal} into \eqref{eq:dEdx}, we arrive at
    \begin{align}\label{eq:boundeq}
        \frac{dE}{dx}\Big\vert_{th} = \frac{2 E_{i}}{m_{\chi}} \left(\sum_{A \subset D}\frac{\mu^{2}_{\chi A}}{m_{A}}n_{A}\sigma_{\chi A}\right)\exp\left[\frac{-2}{m_\chi} \left( x_O \sum_{A\subset O}n_{A}\frac{\mu^{2}_{\chi A}}{m_{A}}\sigma_{\chi A} +x_D \sum_{A \subset D}n_{A}\frac{\mu^{2}_{\chi A}}{m_{A}}\sigma_{\chi A} \right) \right],
    \end{align}
    where $\frac{dE}{dx}\vert_{th}$ is the energy threshold for a plastic track detector and $E_i = m_\chi v_i^2/2$. This equation requires the DM energy to just exceed the detector threshold during its entire transit through the overburden and detector. Fixing $m_\chi$, detector, and overburden parameters, and solving this equation for $\sigma_{\chi A}$ will yield \emph{two} real solutions for $\sigma_{\chi A}$, for all validly bounded $m_{\chi}$ masses. Along with $m_\chi^{\rm max}$, this defines the region of plastic track detection sensitivity.  
\end{enumerate}

Equation~\eqref{eq:boundeq} deserves some discussion, since it is a departure from prior treatments, which separately estimated the threshold and overburden cross section. As just mentioned, for fixed $m_\chi$ there will be two $\sigma_{\chi A}$ solutions to Eq.~\eqref{eq:boundeq}. The larger cross section solution corresponds to what has been called the overburden cross section, and describes a SIMP that scatters a lot with the overburden, but still deposits sufficient energy to leave etchable plastic tracks because of its relatively large cross section. The lower cross section corresponds to what has been called the threshold cross-section, and describes a SIMP that scatters less with the overburden, and in so doing has enough energy to trigger the detector during its passage. What is important, is that both of these cross-sections depend on the dark matter scattering enough with the detector to leave a visible signature, which is encapsulated by Eq.~\eqref{eq:boundeq}, whether SIMP detection is limited by excessive overburden scattering or a small cross-section. We will see that even in the latter case, overburden scattering can still appreciably affect the cross section bound. In the next section we will use this formula to set bounds on dark matter using Skylab and Ohya.

\section{Dark Matter at Skylab and Ohya} 
\label{sec:bounds}

We now use the procedure outlined in Section \ref{sec:dm} to derive bounds from plastic track particle searches at Skylab and Ohya. For each experiment we will describe the number of events $N_c$ that cannot be ruled out as SIMPs, threshold energies $\frac{dE}{dx}\vert_{th}$, along with detector and overburden material. Below we also include a brief description of the etching process in both experiments. Table \ref{Tab:Table1} provides a summary of all essential experimental parameters. For further details about the etching process, see Appendix \ref{app:etch}.\\

\begin{figure}[t!]
    \centering
    \includegraphics[width=0.8\textwidth]{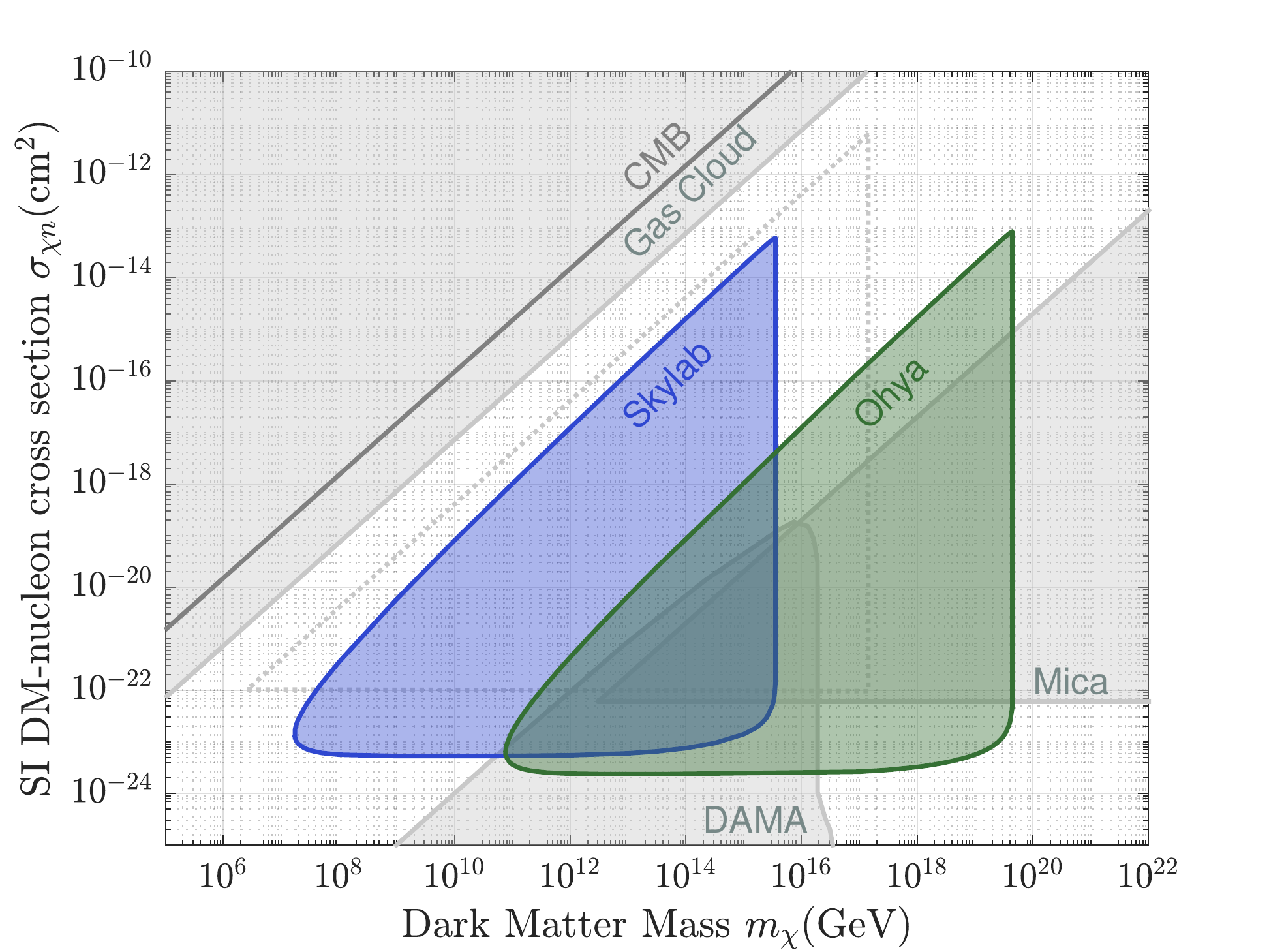}
      \caption{Skylab and Ohya bounds are obtained for a spin-independent dark matter per-nucleon scattering cross section $\sigma_{\chi n}$, using a DM halo velocity distribution, accounting for background events, and other refinements detailed in the text. The prior Skylab bound \cite{Starkman:1990nj}, here reproduced from \cite{Wandelt:2000ad}, is shown with a dotted line. The Mica bound is from \cite{Price:1986ky,Bramante:2018tos}. The DAMA bound is from \cite{Bernabei:1999ui}. We also show the CMB distortion~\cite{Dvorkin:2013cea,Gluscevic:2017ywp} and gas cloud heating~\cite{Bhoonah:2018gjb} bounds from dark matter scattering in the early Universe and in cold gas clouds. For additional complementary bounds at lower cross-sections from XENON1T and MAJORANA, see \cite{Clark:2020mna}.}
         \label{fig:sigmaxn}
\end{figure}

\begin{figure}[t!]
    \centering
    \includegraphics[width=0.8\textwidth]{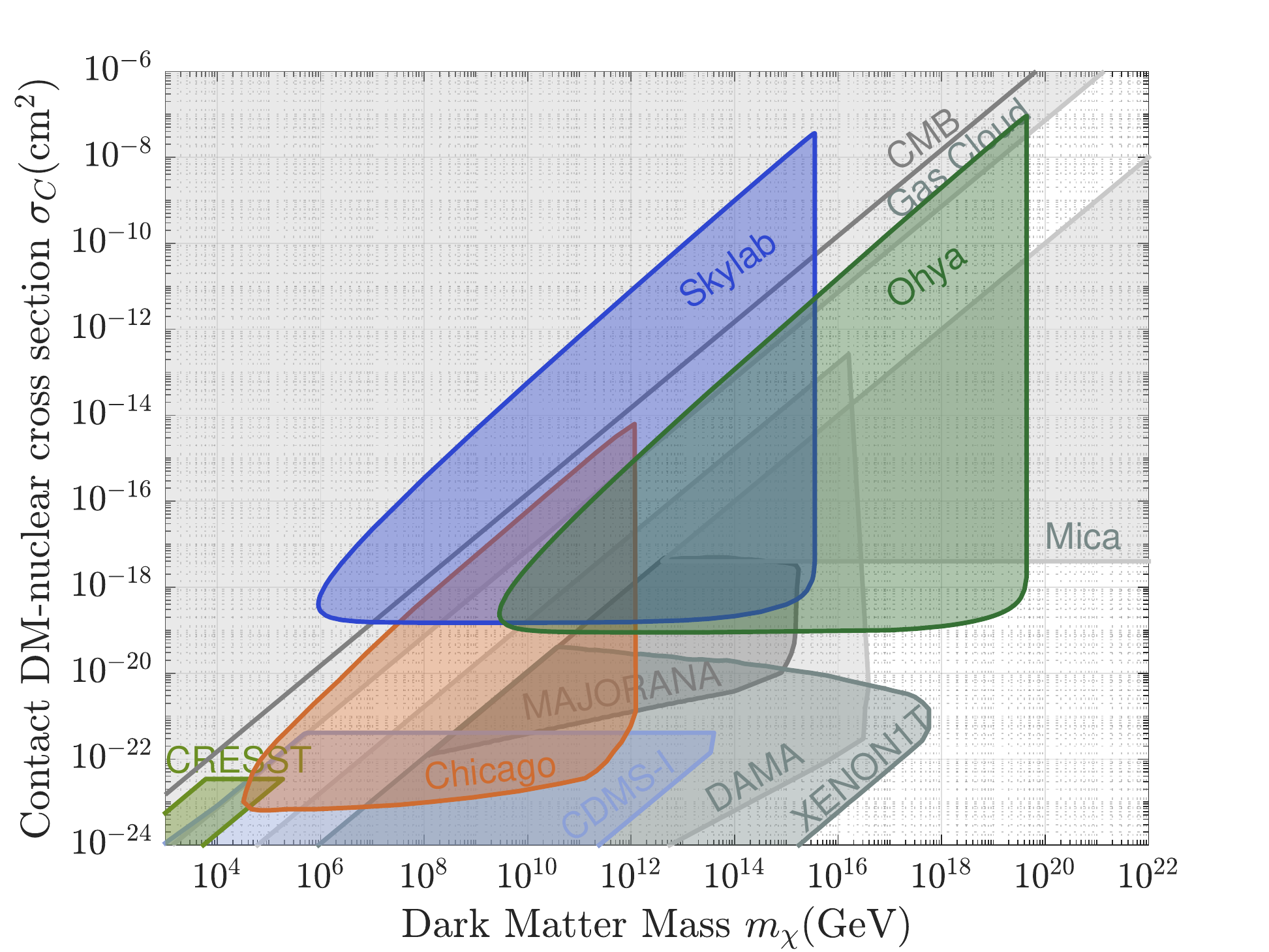}
    \caption{Skylab and Ohya bounds are obtained for a fixed contact DM scattering cross section for all nuclei $\sigma_{c}$ \eqref{eq:sc}, for DM mass $m_\chi$. The Chicago bound is taken from Ref.~\cite{Cappiello:2020lbk} and the Mica bound from Ref.~\cite{Bhoonah:2020dzs}. The limits on contact interactions from DAMA \cite{Bernabei:1999ui} were derived in \cite{Bhoonah:2020dzs}, using a conversion between spin-independent and contact interactions for the lower limit, and an overburden analysis for the upper limit. The CDMS-I and CRESST bound was obtained with a modified version of the code \texttt{verne}~\cite{verne,Kavanagh:2017cru} which accounts for the dark matter flux and overburden \cite{Bhoonah:2020dzs}; recently reported bounds from XENON1T and MAJORANA are also indicated \cite{Clark:2020mna}. We also show the CMB distortion~\cite{Dvorkin:2013cea,Gluscevic:2017ywp} and gas cloud heating~\cite{Bhoonah:2020dzs} constraints arising from dark matter nuclear scattering in the early Universe or in cold gas clouds.}
    \label{fig:sigmaC}
\end{figure}

The \emph{Skylab} plastic track search was conducted on board the Skylab space station, primarily to study the high $Z$ composition of cosmic rays \cite{Skylab}. The detector was a 1.17 m$^{2}$ array of $36$ modules of Lexan plastic track detectors sheets kept onboard for an observation time of $2.19\times 10^{7}$ s. Each Lexan module contained thirty-two stacked $250$ $\mu$m thick Lexan sheets. Assuming uniform stacks laid on top of each other, this amounted to a total of 0.80 cm of Lexan, which has the following composition by weight \cite{Lexan} 
\begin{align}
    \{ \rm H, C, O\}& = \{0.0555, 0.756, 0.189\}. &~~({\rm Lexan}) \nonumber
\end{align}
The overburden for Skylab was a $0.37$ cm aluminum wall to which the Lexan modules were mounted. We conservatively assume that all Skylab events entered the aluminum and Lexan stack at the $\theta_D = 60\degree$ maximum acceptance angle, and hence took the longest possible path through both the overburden material and detector. For a particle entering at a zenith angle of $60\degree$ this corresponds to a distance of $0.74$ cm through the aluminum and 1.6 cm through the Lexan. After etching, \cite{Skylab} identified 150 events in the zenith angle range $\theta_D \leq 60\degree$ which passed largely undeflected through the Lexan sheets. While a future analysis might be able to conclusively attribute these events to high-Z cosmic ray fluxes, here we will treat them as unreduced background events ($N_c$). 

The detector threshold at Skylab was determined using fast ion calibration. Matching the Skylab high-Z ion-scattering threshold using the Bethe formula yields $\frac{dE}{dx}\vert_{th} \approx 0.5 \ $ GeV/cm.

{\em Skylab Etching.} For Skylab, starting from sheet number two, every fourth sheet in each module (except one which was used for calibration purposes) were etched for 160 hours at 40$\degree$C in NaOH solution to which 0.057$\%$ Dowfax surfactant had been added. If a track was observed at coincident sites in multiple sheets, it was assumed that the particle passed through all the sheets in between. Around 150 events were observed that passed through all etched sheets.\\

\begin{table}[t!]
\begin{center}
\def\arraystretch{1.3}
\begin{tabular}{ |c|c|c|} 
\hline
  & \vtop{\hbox{\textbf{Skylab}}} & \textbf{Ohya} \\
\hline
Area A& 1.17 $m^{2}$ & 2442 $m^{2}$ \\
\hline
Duration t& 0.70 yr & 2.1 yr \\
\hline
\hline
Zenith cutoff angle & $\theta_D= 60\degree$ & $\theta_D= 18.4\degree$ \\
\hline
Detector material & \vtop{\hbox{\strut 0.25 mm thick Lexan}\hbox{\strut \ \ \ \ \ \ $\times$ 32 sheets}} & \vtop{\hbox{\strut 1.59 mm thick CR-39}\hbox{\strut \ \ \ \ \ \ \ \ $\times$ 4 sheets }}  \\
\hline
Detector density  & 1.2 g cm$^{-3}$ Lexan & 1.3 g cm$^{-3}$ CR-39 \\
\hline
Detector length at $\theta_D$ & 1.6 cm & 0.66 cm \\
\hline
Overburden density  & 2.7 g cm$^{-3}$ Aluminum & 2.7 g cm$^{-3}$ Rock \\
\hline
Overburden length at $\theta_D$ & 0.74 cm & 39 m \\
\hline
\end{tabular}
\captionof{table}{Summary of the two planar etched plastic track experiments considered in this work, Skylab and Ohya. Along with material elemental compositions given in the text, the quantities shown are sufficient for setting bounds on dark matter interactions using Eqs.~\eqref{eq:maxmass} and \eqref{eq:boundeq}. $\theta_D$ is the maximum zenith angle of acceptance. For Ohya the value of $\frac{dE}{dx} \vert_{th} = 0.3$ GeV/cm corresponds to the 10$\%$ efficiency threshold for nuclearites in Figure 2 of \cite{Orito:1990ny}.}
\label{Tab:Table1}
\end{center}
\end{table}

The \emph{Ohya} experiment was undertaken to search for supermassive relics, nuclearites, and other exotic particles using a 2442 m$^{2}$ array of CR-39 plastic track detectors, consisting of modules of CR-39 placed underground in three caverns located at the Ohya stone quarries in Japan for 767 days \cite{Orito:1990ny}. Each CR-39 module contained four stacked $1.59$ mm thick CR-39 sheets. This amounted to a total of 0.64 cm of CR-39 with the chemical formula $C_{12}H_{18}O_{7}$ indicating the following composition by weight 
\begin{align}
    \{ \rm H, C, O\} = \{0.066, 0.526, 0.408\}. &~~~~(\text{CR-39}) \nonumber
\end{align}
The overburden for Ohya was reported as $10^4$ g cm$^{-2}$ of rock. Assuming a rock density of $\sim$2.7 g\,cm$^{-3}$, this corresponds to 37 meters of rock. For rock's elemental fractions by weight, we use standard values for Earth's crust \cite{mason1982principles}, 
\begin{align}
    \{ \rm O, Si, Al, Fe, Ca, Na, K, Mg, Ti, P \} = \{0.48, 0.28, 0.08, 0.05, 0.03, 0.03, 0.03, 0.02, 0.004 \}. \nonumber ~({\rm Rock}) 
\end{align}
We conservatively assume that all Ohya events entered the rock and CR-39 stack at a $\theta_D = 18.4\degree$ maximum acceptance angle, and hence took the longest possible path through both the overburden material and the detector. 
A particle inbound along this zenith angle traverses 39 m of rock and etches 0.66 cm of CR-39.
After etching, 94 coincident holes were identified, of which 74 were ruled out as manufacturing defects due to their irregular shapes. We therefore consider the remaining 20 holes as candidate dark matter events ($N_c$). 
We note that \cite{Orito:1990ny} determined that only a few of these 20 events could be cosmogenic based on the relative angles of the etched holes.
We nevertheless assume $N_c=20$ candidate events in placing dark matter scattering limits to remain conservative. 

The threshold energy deposition at Ohya was explicitly reported for DM particles that scatter elastically with nuclei. This threshold was given in terms of isotropic flux acceptance, where less flux was accepted for lower thresholds, attained when particles impact the detector from a normal angle
(see Appendix \ref{app:etch}). Here, in order to maximize the accuracy of our overburden modeling and remain conservative, we will use the Ohya 10$\%$ isotropic flux detection efficiency, which implies the aforementioned zenith cutoff angle $\theta_D = 18.4\degree$. The threshold for detection inside this zenith angle was reported to be $\frac{dE}{dx}\vert_{th} = 0.3 \ {\rm GeV ~cm^{-1}}$. 

\textit{Ohya etching.} For Ohya, the etching process involved treating the first two sheets in each module using a NaOH at a temperature of 90$\degree$ C. For 19$\%$ of these, one or two of the sheets were either over-etched or broken and replaced by the corresponding third and/or fourth layers for the etching. The etching process at \cite{Orito:1990ny} was slightly different than \cite{Skylab} in that \cite{Orito:1990ny} required the etching process be strong enough to fuse two adjacent cones from an etch pit into a hole.\\

The important quantities required to set DM bounds using Skylab and Ohya data are summarized in Table~\ref{Tab:Table1}. We note that, as discussed following Eq.~\eqref{eq:flux}, we only consider the downward-going DM flux when setting bounds, since both Ohya and Skylab only provide information about the material directly above their detectors. Figures \ref{fig:sigmaxn} and \ref{fig:sigmaC} show bounds obtained for both spin-independent and contact dark matter scattering on nuclei.

A few features are worth noting in Figures \ref{fig:sigmaxn} and \ref{fig:sigmaC}.
First, the incorporation of a realistic MB velocity distribution for dark matter improves the lower DM cross-section sensitivity of these experiments by an order of magnitude (compared to assuming a fixed DM velocity $\sim 0.001{\rm c}$). This is because our analysis has accounted for the high velocity component of the DM distribution, although we stress that care has been taken not to overstate this component: see discussion following Eq.~\eqref{eq:BoltzDist} and preceding Eq.~\eqref{eq:boundeq}. Secondly, we note that the mass reach of Skylab is substantially reduced as compared to prior estimates appearing in the literature. This is partly because we have accounted for the $\sim 150$ events observed at Skylab as background events that cannot be ruled out as SIMPs, and partly because we have properly required more dark matter's low velocity component to trigger the detector, as the flux becomes more limited at high DM masses (rather than assuming that when the bound is becoming flux-limited, all DM particles will still have $v \sim 0.001{\rm c}$). Finally, we note that requiring the DM maintain enough energy to trigger Skylab during its entire transit through the detector, and using a MB distribution when deriving the bound, has lessened the upper cross section reach by a small amount and altered its scaling with dark matter mass.

\section{Conclusions} 
\label{sec:conc}

This work has provided a detailed derivation of constraints on dark matter's nuclear interactions in etched plastic track detectors, located on board the Skylab space station and forty meters underground in the Ohya quarry. While as compared to previous estimates, the dark matter mass reach of these experiments has been reduced by a few orders of magnitude, the lower cross section sensitivity has improved by an order of magnitude for some dark matter masses.

Both Skylab and Ohya continue to place leading limits on heavy SIMP dark matter's interactions more than three decades after they collected data. In both cases these experiments appear to have been primarily motivated by searches for non-DM particles (cosmic rays and monopoles). Since there exist many dark matter models ($e.g.$ \cite{Nussinov:1985xr,Bagnasco:1993st,Alves:2009nf,Kribs:2009fy,Lee:2013bua,Krnjaic:2014xza,Detmold:2014qqa,Jacobs:2014yca,Wise:2014jva,Wise:2014ola,Hardy:2014mqa,Hardy:2015boa,Gresham:2017zqi,Gresham:2017cvl,Gresham:2018anj,Bramante:2018tos,Ibe:2018juk,Grabowska:2018lnd,Coskuner:2018are,Bai:2018dxf,Bai:2019ogh,Bramante:2019yss,Acevedo:2020avd}) that could have large cross sections and masses, and given the uncharted interaction space evident at high massses in Figures \ref{fig:sigmaxn} and \ref{fig:sigmaC}, it may be prudent to pursue a purpose-built high mass dark matter search program based on plastic track detectors, located both above and below ground, like Skylab and Ohya.

\section*{Acknowledgements}
We thank Nirmal Raj and Aaron Vincent for useful discussions. The work of AB, JB, BC, and NS is supported by the Natural Sciences and Engineering Research Council of Canada (NSERC). Research at Perimeter Institute is supported in part by the Government of Canada through the Department of Innovation, Science and Economic Development Canada and by the Province of Ontario through the Ministry of Colleges and Universities. 

\appendix
\section{Acid etched plastic track detectors} 
\label{app:etch}
Plastic track detectors consist of sheets of polymers that can be damaged by incident fluxes of cosmic rays and other particles, which interact with nuclei and electrons in the material, breaking the bonds binding the constituent monomers together, causing localized damage. The damaged plastic is then cured by immersing it in an acid etching solution. Particle detection is possible, since the curation rate will not be the same across the entire plastic sheet; at the damaged sites the track etch rate, $v_{t}$, is different from the bulk etch rate $v_{b}$, which is the rate at the undamaged sites. The etching process leaves behind conically shaped damage pits. One can either scan adjacent plastic sheets for coincident pairs of conical pits, as in \cite{Skylab}, or alternatively, if the acid etching process is rigorous enough, adjacent pits can merge and form holes that are later observed with a microscope - this was the procedure adopted in \cite{Orito:1990ny}. Neither method of analysis is without complication. Care must be taken to avoid misidentifying manufacturing defects, which are often irregularly shaped as compared to signal events, which have either conical or spherical shapes depending on the etch method. Coincident etched sites across multiple plastic sheets are needed to establish a particle event. Because of these difficulties, the identification process has an efficiency that depends on the size and orientation of the etched hole.

As one might anticipate, the rate of acid etching where particle damage has occurred ($v_{t}$) depends on the energy deposited by the cosmic ray at a damage site. This relationship can be obtained through calibration with a beam of ions, typically iron \cite{Skylab}. The process involves firstly relating $v_{t}$, obtained by dividing the measured size of etch pits by the total etch time \cite{Skylab}, to the effective charge and velocity of the incident beam of charged particles through a power law phenomenological formula. This can then be related to the energy transfer per unit length, $\frac{dE}{dx}$, using $e.g.$ the Bethe-Bloch formula.

The plastic etch method has a threshold, since the process will not identify etch pits below a certain size, implying that the damage done at a site needs to exceed some amount. This sets the minimum amount of energy transfer for detection. For etch holes to form after curation, $v_{t}$ must be greater than $v_{b}$. The exact criterion is given by an expression that depends on the zenith angle $\theta_c$ along which the particle passes through the sheet, since the acid will not etch oblique damage paths as efficiently as for damage paths that are normal to the plastic sheet \cite{Orito:1990ny},
\begin{align}
    \theta_{c} = \cos^{-1}\left(\frac{v_{b}}{v_{t}}\frac{D}{D-d}\right).
\end{align}
Here $D$ and $d$ are the thickness of the plastic track before and after the etching process respectively and $\theta_{c}$ defines a minimum zenith angle below which damaged pits cannot be observed. We note in passing that in the limit $d \ll D$ this is the same requirement set by Shirk and Price in \cite{Skylab}: that the horizontal component of the track etch rate equals the bulk etch rate.

\bibliographystyle{JHEP.bst}

\bibliography{Plastic.bib}

\end{document}